\documentclass[a4paper,preprintnumbers,showpacs,twocolumn,superscriptaddress,nofootinbib,amsmath,amssymb]{revtex4-1}
\usepackage[dvips]{graphics}
\usepackage[hypertex]{hyperref}
\usepackage{color,hyperref}
\usepackage{times}
\usepackage{mathrsfs}
\hypersetup{
  colorlinks=true,        % false: boxed links; true: colored links
  linkcolor=blue,         % color of internal links
  citecolor=magenta,      % color of links to bibliography
  filecolor=magenta,      % color of file links
  urlcolor=blue            % color of external links
}

\def\beq{\begin{equation}}
\def\eeq{\end{equation}}
\def\bear{\begin{eqnarray}}
\def\ear{\end{eqnarray}}

\def\L{\mathscr{L}}

\usepackage{enumitem}
\usepackage{graphicx,subfigure}% Include figure files
\usepackage{dcolumn}% Align table columns on decimal point
\usepackage{bm}% bold math
\usepackage{color}

\begin{document}

\title{Comment on ``Construction of regular
black holes in general relativity"}

\author{Bobir Toshmatov}
\email{bobir.toshmatov@fpf.slu.cz}
\affiliation{Institute of Physics and Research Centre of Theoretical Physics and Astrophysics, Faculty of Philosophy \& Science, Silesian University in Opava, Bezru\v{c}ovo n\'{a}m\v{e}st\'{i} 13,  CZ-74601 Opava, Czech Republic}
\affiliation{Ulugh Beg Astronomical Institute, Astronomicheskaya 33, Tashkent 100052, Uzbekistan}

\author{Zden\v{e}k Stuchl\'{i}k}
\email{zdenek.stuchlik@fpf.slu.cz}
\affiliation{Institute of Physics and Research Centre of Theoretical Physics and Astrophysics, Faculty of Philosophy \& Science, Silesian University in Opava, Bezru\v{c}ovo n\'{a}m\v{e}st\'{i} 13,  CZ-74601 Opava, Czech Republic}

\author{Bobomurat Ahmedov}
\email{ahmedov@astrin.uz}
\affiliation{Ulugh Beg Astronomical Institute, Astronomicheskaya 33, Tashkent 100052, Uzbekistan}
\affiliation{National University of Uzbekistan, Tashkent 100174, Uzbekistan}

\begin{abstract}

We claim that the physical parameters of the constructed black hole solutions in general relativity (GR) coupled to nonlinear electrodynamics (NED) by Zhong-Ying Fan and Xiaobao Wang in the Paper~[Phys. Rev. D 94, 124027 (2016)] are misinterpreted, despite the formalism being correct. We argue that because of these misinterpretations, the derived black hole solutions and the Lagrangian densities presented in that paper are slightly  inconsistent. In this comment, we present complete black hole solutions of the given Lagrangian densities which correct the interpretation of the physical parameters of the constructed black hole solutions and lead to the correct treatment and weak field limits of the physical parameters of the constructed solutions.

\end{abstract}

\maketitle

In the recent paper~\cite{Fan:PRD:2016} (Hereafter, we call this paper as FW16.) a general procedure for constructing exact black hole solutions with the electric or magnetic charges in GR coupled to NED based on the initially interesting and useful proposal of Bronnikov~\cite{Bronnikov2001} has been developed. Here we argue that the Lagrangian densities and the corresponding black hole solutions presented in the FW16 have some discrepancy, i.e., the Lagrangian densities do not correspond to the black hole solutions, or vice versa, because of the misinterpretation of the physical parameters of the constructed black hole solutions.

Here we show that the physical parameters of the obtained solutions in the paper FW16 are misinterpreted. For the more details we direct readers to FW16, without repeating the results presented there.

The action of GR coupled to the NED is given as
\begin{eqnarray}\label{action}
S=\frac{1}{16\pi}\int d^4x\sqrt{-g}\left[R-\mathscr{L}(F)\right]\ ,
\end{eqnarray}
where $F\equiv F_{\mu\nu}F^{\mu\nu}$ is the norm of tensor of the electromagnetic field $F_{\mu\nu}$. According to the formalism the Lagrangian density of the magnetically charged NED can be given by the expression
\bear
&&\L=\frac{4m'}{r^2}\ ,\label{Lagrangian-mg1}\\
&&\L_F=\frac{r^2(2m'-rm'')}{2Q_m^2}\ ,\label{Lagrangian-mg2}
\ear
where $F=2Q_m^2/r^4$ is the electromagnetic field strength produced by the magnetic charge $Q_m$, $m(r)$ is the mass function. Using this Lagrangian densities the authors of the FW16 have obtained magnetically charged Bardeen-like, Hayward-like,  and new type being the Maxwellian in the weak field regime, singular black hole solutions with the line element
\bear\label{line-element}
ds^2=-\left(1-\frac{2m(r)}{r}\right)dt^2+\left(1-\frac{2m(r)}{r}\right)^{-1}dr^2+ r^2d\Omega_2^2\ , \nonumber\\
\ear
where $d\Omega_2^2$ is the solid angle.

Finally, they have generalized these black hole solutions choosing the Lagrangian density in the form
\bear\label{lagrangian-ned}
\L=\frac{4\mu}{\alpha}\frac{(\alpha F)^{\frac{\nu+3}{4}}}{\left[1+(\alpha F)^{\frac{\nu}{4}}\right]^{1+\frac{\mu}{\nu}}}\ ,
\ear
where $\mu>0$ is a dimensionless constant which characterizes the strength of nonlinearity of the electrodynamic field, and $\alpha>0$ is constant parameter
which has the units of the length squared; $\alpha$ is introduced into theory through the definition $Q_m=q^2/\sqrt{2\alpha}$. Moreover, $\nu=2,\mu,1$ correspond to the Bardeen-like, Hayward-like and new Maxwellian black hole solutions, respectively.

In the FW16 the authors have found the general mass function for the magnetically charged black holes in the NED which correspond to the Lagrangian density~(\ref{lagrangian-ned}) in the following form:
\bear\label{lapse-function}
m(r)=M+\frac{q^3}{\alpha}\frac{r^{\mu}}{(r^\nu+q^\nu)^{\mu/\nu}}\ ,
\ear
where $q$ is the magnetic charge parameter and $\nu>0$ is the dimensionless constant. $M$ is the integration constant which was wrongly interpreted in FW16 as the gravitational mass, i.e. $M=M_g$. Indeed, if the mass function~(\ref{lapse-function}) is substituted in the Lagrangian density~(\ref{Lagrangian-mg1}), one obtains the Lagrangian density~(\ref{lagrangian-ned}) which seems correct. However, if the Einstein equations or Eq.~(\ref{Lagrangian-mg1}) with the Lagrangian density~(\ref{lagrangian-ned}) are solved with respect to the mass function $m(r)$, one obtains the mass function in the following form:
\bear\label{lapse-function-new}
m(r)=M_g-\frac{q^3}{\alpha}\left[1 -\frac{r^{\mu}}{(r^\nu+q^\nu)^{\mu/\nu}}\right]\ ,
\ear
which with comparison to~(\ref{lapse-function}) contains extra term $q^3/\alpha$ in the right hand side. If one drops this term, all equations are satisfied, however, it is equivalent to the condition $q=0$, which eliminates the last term together with the NED as well. Therefore, the solution~(27) together with the Bardeen-like~(15), Hayward-like~(22), and new type~(24) black hole solutions obtained in the FW16 are incomplete, since $M$ in (\ref{lapse-function}) is stated as gravitational mass.  Or the solution (\ref{lapse-function}) can be considered complete only with wrong interpreted mass parameters. Below we address the interpretation of the correct mass parameters.

Absence of the term $-q^3/\alpha$ in the mass function leads to several physically inappropriate black hole properties which are presented in the FW16. For example, if one studies the Arnowitt-Deser-Misner (ADM) mass, the asymptotic behaviour of~(\ref{lapse-function-new}) gives the ADM mass of the black hole to be $M_{ADM}=M_g$ as for the Reissner-Nordstr\"{o}m black hole in GR coupled to the linear (Maxwell) electrodynamics. However, if one considers the mass function~(\ref{lapse-function}), then, in the asymptotics the ADM mass takes the value $M_{ADM}=M_g+q^3/\alpha$~\cite{Fan:PRD:2016} which is in the contradiction with that of the Reissner-Nordstr\"{o}m black hole. One may argue that nonlinearity of the electromagnetic field leads the additional mass to the the pure gravitational mass, $M_g$, however, even if it is alright, at least in the new type of black hole solution which is Maxwellian in the weak field limit, the ADM mass must be equal to the pure gravitational mass as it has been stated in~\cite{Bronnikov:PRL:2000}. The correctness of the solution of the FW16 can be recovered only if $M_g$ is considered not the pure gravitational mass (Despite, they stated it as pure gravitational mass), instead it is effective mass with the definition $M_{eff}=M_g-q^3/\alpha$.

As shown in the FW16, the black hole solution~(\ref{line-element}) with the mass function~(\ref{lapse-function}) is singular at origin, $r=0$, and regular only if the pure gravitational mass (or so-called Schwarzschild mass) is neglected, $M_g=0$, and $\mu\geq3$. Existence of the regular black hole without gravitational mass is mathematically alright, but it seems somehow pathologic from the physical point of view. In the case of the solution~(\ref{lapse-function-new}), it is also singular at the origin $r=0$,  even if its mass $M_g=0$. The only way to make the black hole regular everywhere of the spacetime is to assume that the gravitational mass is equal to the so-called electromagnetically induced mass $M_{em}$
\bear\label{reg-condition}
M_g=M_{em}\equiv\frac{q^3}{\alpha}\ ,
\ear
with the condition $\mu\geq3$. Then, one can write the metric function~(\ref{lapse-function-new}) in the following form~\footnote{Regular black hole solution derived in the FW16 can also be written in the form of~(\ref{lapse-function-new1}) but instead of the gravitational mass, $M$, induced electromagnetic mass, $M_{em}$ has to be inserted.}:
\bear\label{lapse-function-new1}
f(r)=1-\frac{2M_gr^{\mu-1}}{(r^\nu+q^\nu)^{\frac{\mu}{\nu}}}\ .
\ear
Here $M_g$ is still pure gravitational mass (or electromagnetically induced mass since $M_g=M_{em}$). And now one can construct the Bardeen-like, Hayward-like and new type regular black hole solutions in GR coupled to NED by changing $\nu=2$, $\nu=\mu$, and $\nu=1$, respectively, only if $\mu\geq3$.

Let us discuss the electrically charged asymptotically flat black hole solutions in GR coupled to the NED~(36) derived in the FW16. Indeed, as in the case of the magnetically charged black hole solution, the method of obtaining the solution is correct and well explained. The Lagrangian density is given by
\bear
\L=\frac{2m''}{r}\ ,\label{Lagrangian-el1}
\ear
with the ansatz
\bear
A=\frac{3m-rm'}{2Q_e}dt\ ,\label{ansatz}
\ear
where $Q_e=q^2/\sqrt{2\alpha}$. Again, the authors aimed to obtain the black hole solution with the mass function~(\ref{lapse-function}). However, the ansatz in Eq.~(36) does not correspond to the mass function~(\ref{lapse-function}), whereas it corresponds to the one in~(\ref{lapse-function-new}). Thus, the ansatz and the metric function are inconsistent. Moreover, the electrically charged black hole solution with the mass function~(\ref{lapse-function}) in Eq.~(36) is not a solution of the Lagrangian density given in~(38) of the FW16. These are also by virtue of the misinterpreted mass parameters in FW16.

One may argue that the correspondence of the black hole solutions to the Lagrangian densities presented in the FW16 is justified by the thermodynamics of them as in the FW16 the authors studied the first law of thermodynamics following to Zhang and Gao~\cite{Zhang} by considering the parameter $\alpha$ as also thermodynamical variable and introducing the new conjugate potential $\Pi$ to $\alpha$ as $dM_{ADM}=TdS+\Phi dQ_e+\Psi dQ_m+\Pi d\alpha$, where $\Phi$ and $\Psi$ are the conjugate potentials to the electric $Q_e$ and magnetic $Q_m$ charges, respectively. Here $\Pi$ is directly related to the Lagrangian density as $\Pi=1/4\int_{r_0}^\infty dr\sqrt{-g}\partial\L/\partial\alpha$. According to the FW16, equality of the left and right hand sides of the Smarr formula ($M_{ADM}=2TS+\Phi Q_e+\Psi Q_m+2\Pi\alpha$) confirmed the correctness of the solution as
\bear\label{smarr}
M_{ADM}=\frac{r_0}{2}+\frac{q^3}{\alpha}\left[1 -\frac{r_0^{\mu}}{(r_0^\nu+q^\nu)^{\mu/\nu}}\right]\ ,
\ear
where $M_{ADM}=M+q^3/\alpha$ for the solution~(\ref{lapse-function}) with gravitational mass $M$ in FW16. However, this is not true, since in all the thermodynamic quantities the mass function~(\ref{lapse-function}) has not been used, otherwise, the relation~(\ref{smarr}) would not be correct. Instead, in all the thermodynamic quantities the correct mass function~(\ref{lapse-function-new}) is used since, for example, although the Hawking temperature $T$ and entropy $S$ are directly found from the metric function (mass function), and have been written in terms of the horizon, $r_0$, inexplicitly, they have the same form as in (46) and (47) of the FW16 for the mass functions~(\ref{lapse-function}) and~(\ref{lapse-function-new}); the charge conjugate potentials, $\Phi$ and $\Psi$, presented in the FW16 correspond to~(\ref{lapse-function-new}), rather to~(\ref{lapse-function}). For the solution with the mass function~(\ref{lapse-function-new}), the Smarr formula gives the same result in~(\ref{smarr}), but here the right hand side of~(\ref{smarr}) is equal to the pure gravitational mass $M$. This proves the correspondence of the black hole solution with mass function~(\ref{lapse-function-new}) to the Lagrangian~(\ref{lagrangian-ned}).

Finally, in the last section of the FW16 where the charged asymptotically anti-de Sitter black hole solution was derived in the NED also above mentioned problems are repeated in the both black hole solution and the thermodynamics as well.

To conclude, to keep the results of the paper FW16 still correct, we suggest the readers of the paper FW16 replacing notion of the pure gravitational mass $M$ (according to the notations in~\cite{Fan:PRD:2016}) in the FW16 with notion of the effective mass $M_{eff}$ which is defined by the difference between pure gravitational mass $M_g$ and electromagnetically induced mass $M_{em}=q^3/\alpha$ as $M_{eff}=M_g-M_{em}$.

In addition, we should note that this comment together with the recent one~\cite{Bronnikov:PRD:2017} can be considered as the refinement to the interesting and relevant paper~\cite{Fan:PRD:2016} which is still very useful contribution to the studies of the black holes in GR coupled to the NED, since one can consider that the derivation of the new generalized Bardeen, Hayward and the new class of (Maxwellian) regular and singular black hole solutions (despite the Lagrangian densities corresponding to the solutions are partially incomplete, or vice versa) as well as generalizing these solutions into one form themselves are the great job. Without the paper~\cite{Fan:PRD:2016}, additional paper~\cite{Bronnikov:PRD:2017} and this paper would have never been appeared. Afterwards, interesting topics were treated later under influence of the FW16, for instance, in the paper~\cite{Toshmatov:PRD:2017} we have derived the possible rotating counterparts of the obtained solutions by FW16. Moreover, in~\cite{Lamy:2018} image of the supermassive black holes in the center of the Galaxy by considering the Sgr $A^*$ is nonsingular rotating black holes derived in~\cite{Toshmatov:PRD:2017}, construction of the new class of regular black hole solutions by using the weak and dominant energy conditions in~\cite{Rodrigues:2018}, the Smarr formula for charged black holes in nonlinear electrodynamics in~\cite{Balart} have been studied. For more -- see~\cite{Fan2017,Nojiri,Rodrigues,TSSA}.

\section*{Acknowledgements}

The authors would like to thank the anonymous referee for useful suggestions which improved quality of the paper. B.T. and Z.S. would like to acknowledge the institutional support of the Faculty of Philosophy and Science of the Silesian University in Opava, the internal student grant of the Silesian University (Grant No.~SGS/14/2016) and the Albert Einstein Centre for Gravitation and Astrophysics under the Czech Science Foundation (Grant No.~14-37086G). The research is partially supported by Grants No.~VA-FA-F-2-008 and No.~YFA-Ftech-2018-8 of the Uzbekistan Ministry for Innovation Development, by the Abdus Salam International Centre for Theoretical Physics through Grant No.~OEA-NT-01 and by Erasmus+ exchange grant between Silesian University in Opava and National University of Uzbekistan.

\label{lastpage}

\end{document}